**Experimental Demonstration of >230° Phase Modulation in Gate-Tunable Graphene-Gold Reconfigurable Mid-Infrared Metasurfaces**

Michelle C. Sherrott[1,2,‡], Philip W.C. Hon[2,3,‡], Katherine T. Fountaine[2,3], Juan C. Garcia[3], Samuel M. Ponti[3], Victor W. Brar[1,4], Luke A. Sweatlock[2,3], Harry A. Atwater[1,2]*

1. Thomas J. Watson Laboratory of Applied Physics, California Institute of Technology, Pasadena, CA 91125, USA
2. Resnick Sustainability Institute, California Institute of Technology, Pasadena, CA 91125, USA
3. Northrop Grumman Corporation, NG Next Nanophotonics & Plasmonics Laboratory, Redondo Beach, CA 90278, USA
4. Department of Physics, University of Wisconsin-Madison, Madison, WI 53706, USA

[‡] Equal contributors

*Corresponding author: Harry A. Atwater (haa@caltech.edu)

**Abstract:**

Metasurfaces offer significant potential to control far-field light propagation through the engineering of amplitude, polarization, and phase at an interface. We report here phase modulation of an electronically reconfigurable metasurface and demonstrate its utility for mid-infrared beam steering. Using a gate-tunable graphene-gold resonator geometry, we demonstrate highly tunable reflected phase at multiple wavelengths and show up to 237° phase modulation range at an operating wavelength of 8.50 μm. We observe a smooth monotonic modulation of phase with applied voltage from 0° to 206° at a wavelength of 8.70 μm. Based on these experimental data, we demonstrate with antenna array calculations an average beam steering efficiency of 50% for reflected light for angles up to 30°, relative to an ideal metasurface, confirming the suitability of this geometry for reconfigurable mid-infrared beam steering devices.

Keywords: Metasurface, graphene, phase modulation, field-effect modulation, beam steering, mid-infrared

Metasurfaces have been demonstrated in recent years to be powerful structures for a number of applications including beam steering[1], focusing/lensing[2, 3], and more complex functionalities such as polarization conversion, cloaking, and three-dimensional image reconstruction[4-8], among others[9-14]. These functionalities are accomplished through careful engineering of phase fronts at the surface of a material, where geometric parameters of resonant structures are designed to scatter light with a desired phase and amplitude. However, all of these structures have functions that are fixed at the point of fabrication, and cannot be transformed in any way. Therefore, significant effort has been made in the community to develop metasurfaces that can be actively modulated. There exist numerous examples of metasurface designs which enable active control of reflected or transmitted amplitude, taking advantage of different technologies including MEMS, field-effect tunability, and phase change materials[15-19], discussed further in recent reviews of the state-of-the-art in metasurfaces[13, 20, 21].

For mid-infrared (mid-IR) light, graphene has been demonstrated as an ideal material for active nanophotonic structures for a number of reasons, including its low losses in the mid-IR and its intermediate carrier concentration ($10^{12}$ - $10^{13}$ cm$^{-2}$), placing its plasma frequency in the IR – THz regime[22-26]. Additionally, since it is atomically thin and has a linear density of electronic states, its charge carrier density can be easily modulated via electrostatic gating in a parallel plate capacitor configuration[27-30]. Its corresponding complex permittivity can therefore be modulated over a wide range, potentially at GHz speeds. Recent works have demonstrated that the incorporation of graphene into resonant gold metasurfaces can also be used to significantly modulate absorption profiles, operating at MHz switching speeds[18]. This has been accomplished by either the un-assisted modulation of the graphene dielectric constant, or by exploiting the strong confinement of light by a graphene plasmon excited between metal edges to enhance the sensitivity of the design to the graphene's optical constants.[17, 31] Additional examples have used the tunable permittivity of graphene to modulate the transmission characteristics of a variety of waveguide geometries[32, 33].

Despite the significant progress that has been made, an important requirement for power efficient, high-speed, active metasurfaces is electrostatic control of scattered phase at multiple wavelengths, which has not been adequately addressed experimentally in the mid-IR. In gaining active control of phase, one can engineer arbitrary phase fronts in both space and time, thereby opening the door to reconfigurable metasurface devices. This is particularly necessary as classic techniques for phase modulation including liquid crystals and acousto-optic modulators are generally poorly-suited for the IR due to parasitic absorption in the materials used[34, 35], in addition to being relatively bulky and energy-expensive in comparison to electrostatic modulators. Similarly, though 60° phase modulation based on a $VO_2$ phase transition has been demonstrated at 10.6 μm, the phase transition occurs over relatively long time scales and the design is limited in application due to the restricted tunability range[36]. Finally, recent works on the electrostatic control of phase in the mid-IR using graphene-integrated or ITO-integrated resonant geometries are limited to only 55° electrostatic phase tunability at 7.7 μm[37] and 180° tunability at 5.95 μm[38], respectively. In this work, we overcome these limitations and experimentally demonstrate widely-tunable phase modulation in excess of 200° with over 250 nm bandwidth using an electrostatically gate-tunable graphene-gold metasurface (see Figure 1). We highlight a smooth phase transition over 206° at 8.70 μm and sharper, but larger, phase modulation of 237° at 8.50 μm, opening up the possibility of designing high efficiency, reconfigurable metasurface devices with nanosecond switching times. By measuring this active tunability over multiple wavelengths in a Michelson interferometer measurement apparatus, we present evidence that this approach is suitable for devices that can operate at multiple wavelengths in the mid-IR.

Our tunable phase metasurface design is based on a metasurface unit cell that supports a gap plasmon mode, also referred to as a patch antenna or 'perfect absorber' mode, which has been investigated previously by many groups[39-42], shown schematically in Figure 1a. Absorption and phase are calculated as a function of Fermi energy ($E_F$) using COMSOL FEM and Lumerical FDTD software (see Methods for calculation details). A 1.2 μm length gold resonator on graphene is coupled to a gold back-plane, separated by 500 nm $SiN_x$. At the appropriate balance of geometric and materials parameters, this structure results in near-unity absorption on resonance, and a phase shift of 2π. This may be considered from a theoretical perspective as the tuning of parameters to satisfy critical coupling to the metasurface[40]. This critical coupling occurs when the resistive and radiative damping modes of the structure are equal, thereby efficiently transforming incoming light to resistive losses and suppressing reflection. This condition is possible at subwavelength spacing between the gold dipole resonator and back-plane, when the resonator is able to couple to its image dipole moment in the back-reflector, generating a strong magnetic moment. The magnetic moment, in turn, produces scattered fields that are out of phase with the light reflected from the ground plane, leading to destructive interference and total absorption. This may be considered the plasmonic equivalent of the patch antenna mode.

In order to enhance the sensitivity of the structure to the tunable permittivity of the graphene, these unit cells are arrayed together with a small (50 nm) gap size to result in significant field enhancement at the position of the graphene, as shown in Figures 1b and 1c. This is critical for enhancing the in-plane component of the electric field to result in sensitivity to the graphene's optical constants. Therefore, as the Fermi energy of the graphene is modulated, changing both the inter- and intra-band contributions to its complex permittivity, the resonant peak position and amplitude are shifted, as shown in Figures 1d and 1e. Specifically, the intraband contribution to the permittivity is shifted to higher energies as the plasma frequency of the graphene, $\omega_p$ increases with the charge carrier density as $\omega_p \propto n^{1/4}$. Additionally, as $E_F$ increases, Pauli blocking prevents the excitation of interband transitions to energies above $2E_F$, thereby shifting these transitions to higher energy. The net effect of these two contributions is a decrease of the graphene permittivity with increasing carrier density, leading to a shift of the gap mode resonance to higher energy.

By taking advantage of graphene's tunable optical response, we obtain an optimized design capable of a continuously shifted resonance peak from 8.81 μm at $E_F$ = 0 eV or Charge Neutral Point (CNP) to 8.24 μm at $E_F$ = 0.5eV; a peak shift range of 570 nm. Correspondingly, this peak shift indicates that at a fixed operation wavelength of 8.50 μm, the scattered phase can be modulated by 225°, as seen in Figure 1f. This trend persists at longer wavelengths, with greater than 180° modulation achieved between 8.50 and 8.75 μm. At shorter wavelengths, such as 8.20 μm, minimal tuning is observed because this falls outside of the tuning range of the resonance. It is noteworthy that this phase transition occurs sharply as a function of $E_F$ at 8.50 μm because it falls in the middle of our tuning range, and becomes smoother at longer wavelengths. We therefore illustrate this smooth resonance detuning at a wavelength of 8.70 μm in Figure 1b and 1c, wherein we plot the magnitude of the electric field at different Fermi energies of the graphene. On resonance (Figure 1b), the field is strongly localized to the gap, and then as the Fermi energy is increased (Figure 1c), this localization decreases as the gap mode shifts to shorter wavelengths. Field profiles at 8.50 μm are presented in Supporting Information Section I. These different responses are summarized at three wavelengths (8.2, 8.5, and 8.7 μm) in Figure 1f, where the phase response is plotted as a function of $E_F$.

We experimentally demonstrate the tunable absorption and phase of our designed structure using Fourier-Transform Infrared Microscopy and a mid-IR Michelson interferometer, respectively, schematically shown in Figure 2a. Graphene-gold antenna arrays are fabricated on a 500 nm free-standing SiNx membrane with a gold back-plane. A Scanning Electron Microscope (SEM) image of the resonator arrays is presented in Figure 2b. An electrostatic gate voltage is applied between the graphene and gold reflector via the doped silicon frame to modulate the Fermi energy. Tunable absorption results are presented in Figure 2c demonstrating 490 nm of tunability from a resonance peak of 8.63 μm at the CNP of the graphene to 8.14 μm at $E_F$ = 0.42 eV, corresponding to voltages of +90 V and -80V. This blue-shifting is consistent with the decrease in graphene permittivity with increasing carrier concentration, and agrees well with simulation predictions. Discrepancies between simulation and experiment are explained by fabrication imperfections, as well as inhomogeneous graphene quality and minor hysteretic effects in the gate-modulation due to the $SiN_x$ and atmospheric impurities[43]. The minor feature at 7.6 μm is an atmospheric artifact. The shoulder noted especially at longer wavelengths is a result of the angular spread of the FTIR beam, wherein the use of a 15X Cassegrain objective results in off-normal illumination of the sample, explained further in Supporting Information Section II. We note that the processing of our sample in combination with the surface charge accumulated as a result of the $SiN_x$ surface results in a significant hole-doping of the graphene, as has been observed in previous experiments[44]. Due to this heavy doping, we are unable to experimentally observe the exact CNP of the graphene using standard gate-dependent transport measurement techniques, and therefore determine this by comparison to simulation. We then calculate the Fermi energy at each voltage using a standard parallel plate capacitor model.

To experimentally characterize the phase modulation of scattered light achievable in our graphene-gold resonant structure, we use a custom-built mid-IR, free-space Michelson Interferometer, for which a schematic is presented in Figure 3a and explained in depth in the Methods section. The integrated quantum cascade laser source, MIRcat, from Daylight Solutions provides an operating wavelength range from 6.9 μm to 8.8 μm, allowing us to characterize the phase modulation from our metasurface at multiple wavelengths. The reference and sample legs of the interferometer have independent automated translations in order to collect interferograms at each wavelength as a function of gate voltage.

A comparison of the relative phase difference between interferograms taken for different sample biases is conducted to capture the phase shift as a function of $E_F$. At each Fermi energy, an interferogram for different reference mirror displacements is taken. Due to the different absorptivity at each doping level, each biases' interferogram is normalized to its own peak value. We then take the midpoint of the normalized interferogram amplitude as a reference, and a relative phase shift from one bias to the other is calculated by recording the displacement between the two interferograms at the reference amplitude. Factoring that the sample leg is an optical double pass, the relative phase difference is given by equation 1:

$$\Delta\phi = \frac{720\Delta x}{\lambda} \qquad \text{(Equation 1)}$$

Where $\Delta\Phi$ is the phase difference between different sample responses in degrees, $\Delta x$ is the displacement between interferograms, and $\lambda$ is the wavelength of operation. Data collected for three Fermi energies at 8.70 μm and fitted to a linear regression for extracting phase based on the above equation are presented in Figure 3b. For straightforward comparison, the phase modulation is presented relative to zero phase difference at $E_F = 0$ eV. Linear regression fits to the data for all Fermi energies measured at 8.70 μm are presented in Figure 3c, and the extracted phase as a function of $E_F$ is presented in Figure 3d. Discrepancies between the experimental data and fits, particularly at CNP, can be explained by the decreased reflection signal from the sample due its strong absorption on resonance.

To further highlight the broad utility of our device, phase modulation results are presented in Figure 4a at multiple wavelengths: 8.20, 8.50, and 8.70 μm. At an operating wavelength of 8.70 μm, continuous control of phase is achieved from 0° relative at CNP to 206° at $E_F$ = 0.44 eV with excellent agreement to simulation. At 8.50 μm, this range increases to 237°, much greater than any observed in this wavelength range previously, though as noted above, the transition is very sharp. At the shorter wavelength of 8.20 μm, a modulation range of 38° is achieved, with excellent agreement to simulation, demonstrating the different trends in phase control this structure presents at different wavelengths. Simulation parameters are presented in the Methods section. Deviation is primarily due to hysteresis effects and sample inhomogeneity. We summarize the experimental and simulation results at all wavelengths between 8.15 μm and 8.75 μm in Figure 4b, wherein we plot the tuning range at each wavelength, defined as the maximum difference of scattered phase between CNP and $E_F$ = 0.44 eV. This Fermi energy range is limited by electrostatic breakdown of the $SiN_x$ gate dielectric. The data from wavelengths not presented in Figure 4a are included in Supporting Information Section III. We can therefore highlight two features of this structure: at longer wavelengths, we observe experimentally a smooth transition of phase over more than 200°, and at slightly shorter wavelengths, we can accomplish a very large phase tuning range with the tradeoff of a large transition slope. It is also noteworthy that more than 200° active tunability is achieved between 8.50 μm and 8.75 μm, which is sufficient for active metasurface devices in the entire wavelength range.

To illustrate the applicability of our design to reconfigurable metasurfaces, we calculate the efficiency of beam steering to different reflected angles as a function of active phase range for a linear array of independently gate-tunable elements as shown schematically in Figure 5a. We choose a linear array with polarization orthogonal to the steering direction to ensure minimal coupling between neighboring elements and a pitch of 5.50 μm to suppress spurious diffracted orders at a wavelengths of 8.50 μm. To quantify the beam steering feasibility of this metasurface, we frame the analysis in the formalism of antenna array theory, where the array can be considered as a discretized aperture. The far-field radiation pattern of such a discretized aperture can be analytically calculated by independently considering the physical array configuration (radiating element layout) and the radiating element properties, such as its amplitude, phase and element far-field radiation pattern. For a general two dimensional array, the far-field radiation pattern is given by the array factor weighted by the element's radiation pattern. The element pattern can be considered a weighting factor in the calculation of the far-field radiation pattern, where the array factor is only a function of the element placement and assumed isotropic radiators with a complex amplitude and phase. For relatively omnidirectional radiating elements, as in our case, the array factor captures the primary radiation pattern features, such as the main beam direction, main beam half power beam width (angular width of the main beam noted at half the main beam peak intensity), and major side lobes, reasonably well. The array factor for a general two-dimensional configuration is given as:[45]

$$AF(\theta, \varphi) = \sum_{n=1}^{N} \sum_{m=1}^{M} I_{mn} e^{j\alpha_{mn}} e^{j\gamma_{mn}} \qquad (1)$$

$$\alpha_{mn} = -\beta[x'_{mn} \sin\theta_0 \cos\varphi_0 + y'_{mn} \sin\theta_0 \sin\varphi_0] \qquad (2)$$

$$\gamma_{mn} = \beta \hat{r} \cdot \hat{r}'_{mn} = \beta[x'_{mn} \sin\theta \cos\varphi + y'_{mn} \sin\theta \sin\varphi] \qquad (3)$$

where $\theta_0$ and $\varphi_0$ are the elevation and azimuthal values of the main beam pointing direction, respectively, $\alpha_{mn}$ represents the element imparted phase that controls the beam direction, $\gamma_{mn}$ represents the path length phase difference due to the element position $\hat{r}'_{mn}$ and the unit vector $\hat{r}$ from the array center to an observation angle, $\theta, \varphi$. $\beta$ is the free space propagation constant, $I_{mn}$ is the complex element amplitude and the double summations represent the row and column element placement of a general two-dimensional array.

Considering only the array factor, we can analytically capture the beam steering characteristics of a metasurface as a function of the achievable element phase tuning range. In the microwave regime, where the achievable element phase tuning range is greater than 270°, beam attributes such as its pointing direction and side lobe levels, can be quantified as a function of the phase discretization; the phenomenon is known as quantization error[46]. Independent of quantization errors, it is informative to understand the consequence of an element phase tuning range well below the desired ideal 360°. We define a figure of merit, the beam efficiency η, to be the ratio of the power in the half power beam width for a limited phase tuning range relative to the total radiated power steered to a given angle for an ideal case (tuning range of 360°). In our analysis we consider maximum phase tuning ranges as low as 200° and desired scan angles up to +/- 30° relative to surface normal. For phase ranges below 200°, the undesirable side lobes will equal or exceed the intensity of the primary beam and main beam pointing errors exceeding one degree can exists; therefore, we restrict our analysis for phase ranges greater than 200°. In a simplified analysis, a one-dimensional array is assumed (Fig. 5a). Since the focus of the analysis is only on the consequence of a limited element phase tuning range, the element amplitude is assumed to be equal and unity. Assuming a fine enough gating step size, a virtually continuous sampling of a given element phase tuning range, is possible and therefore quantization error is not an issue. In this analysis, for a calculated element phase value that was unachievable, the closest phase value achievable was assigned. Namely, either an element phase value of 0° or the maximum phase for the considered element phase tuning range. As shown in Figure 5b, regardless of the element phase tuning range, the main beam scanning direction of zero degrees represents the trivial case where a zero difference in beam efficiency is expected because all elements exhibit the same reflected phase (zero phase gradient along the metasurface). The analysis illustrates the trade space and suggests, for the experimentally verified 237° of element phase tuning at 8.50 μm, that beam steering efficiency is on average 50% up to a scan angle of ±30°. Below this, lower efficiency steering is observed; however, we note that down to 200°, the steered main beam signal still exceeds the intensity in the other lobes. We note that the fluctuating trends observed as a function of reflection angle are a result of the incomplete phase range, which manifests differently depending on the deviation from the ideal phase gradient needed.

This clearly illustrates the necessity of achieving at least 200° in active phase control in order to create viable reconfigurable metasurfaces. In addition, it is noteworthy that this calculation includes an assumption of all intermediate phase values being available, meaning that a smoothly varying phase response as a function of gate voltage is necessary, as demonstrated in our device. This highlights the potential applications of our structure to metasurface devices, in which independently gateable elements can be used to generate arbitrary phase gradients in time and space.

In conclusion, we have demonstrated for the first time electrostatic tunability of phase from graphene gold antennas of 237° at a wavelength of 8.5 μm, more than 55° greater than has been demonstrated in the mid-IR in a different materials system. We additionally demonstrate phase modulation at multiple wavelengths, exceeding 200° from 8.50 to 8.75 μm. By calculating from antenna theory the fraction of power reflected to the desired angle as opposed to spurious side-lobes, we show that this design will enable beam steering with acceptable signal to noise ratio. We therefore conclude that this design is feasible for reconfigurable metasurfaces.

**Methods:**

Device Fabrication:

Graphene was grown on 50 μm thick Cu foil using previously established CVD methods[24, 47, 48]. Following growth, the graphene was spin-coated with 2 layers of poly(methyl methacrylate) (PMMA). Cu foil was etched away in iron chloride solution, and the graphene was transferred to a suspended $SiN_x$ membrane obtained commercially from Norcada, part #NX10500E. A back-reflector/back-gate of 2 nm Ti/200 nm Au was evaporated on the back of the membrane by electron beam deposition. 100keV electron beam lithography was then used to fabricate the device. First, arrays of gold resonators were patterned in 300 nm thick 950 PMMA (MicroChem) developed in 3:1 isopropanol:methyl isobutyl ketone (MIBK) for one minute. The sample was then etched for five seconds in a RIE oxygen plasma at 20mTorr and 80W to partially remove the exposed graphene. 3 nm Ti/60 nm Au was then deposited by electron beam evaporation, and liftoff was done in acetone heated to 60°C. A second electron beam lithography step was used to define contacts of 10 nm Ti/150 nm Au. Wire bonding was done to electrically address the electrode.

Electromagnetic Simulations:

We use commercially available finite element methods software (COMSOL) to solve for the two-dimensional complex electromagnetic field of our structures. Graphene is modeled as a thin film of thickness $\delta$ with a relative permittivity from the Kubo formula $\varepsilon_G = 1 + 4\pi i\sigma/\omega\delta\varepsilon_0$. $\sigma(\omega)$ is the complex optical constant of graphene evaluated within the local random phase approximation[23]. The value of $\delta$ is chosen to be 0.1 nm which shows good convergence with respect to a zero-thickness limit. The complex dielectric constant of $SiN_x$ was fit using IR ellipsometry based on the model in ref 49. Three-dimensional simulations are performed using finite difference time domain (FDTD) simulations (Lumerical). Graphene is modeled as a surface conductivity adapted again from ref 23. We use a scattering rate of 20 fs for the graphene, which provides the optimum fit to experimental results and is consistent with previous experimental works using patterned CVD graphene on $SiN_x$.[44]

Interferometry Measurements:

A custom built mid-IR, free-space Michelson interferometer was used to characterize the electrically tunable optical reflection phase from the graphene-gold metasurface. The integrated quantum cascade laser source, MIRcat, from Daylight Solutions provided an operating wavelength range from 6.9 μm to 8.8 μm, which was a sufficiently large enough wavelength range to characterize the absorption spectra and phase of the designed metasurface. A ZnSe lens with a focal length of 75 mm was used to focus the beam onto the sample. The near-field beam waist was 2.5 mm and the far-field beam waist was 90 μm and was measured with a NanoScan beam profiler. The reference and sample legs have independent automated translations, namely, the reference mirror is mounted on a Newport VP-25XA automated linear translation stage with a typical bi-directional repeatability of +/- 50 nm and the sample stage is automated in all three dimensions to give submicron alignment accuracy with the Newport LTA-HS. The propagating beams from the sample and reference legs combine after a two inch Germanium beam splitter. Two ZnSe lenses, one with a focal length of 100 mm and another with 1000 mm image the beam at the sample plane with a ~10 times expansion. Control of the source, translation stages, pyroelectric power detector and the Keithley source used to bias the metasurface is conducted through a Labview automation script.

**Acknowledgements:**
This work was supported by the U.S. Department of Energy (DOE) Office of Science, under Grant No. DE-FG02-07ER46405. M.C.S. acknowledges support by the Resnick Sustainability Institute. This research used facilities of the DOE 'Light-Material Interactions in Energy Conversion' Energy Frontier Research Center.

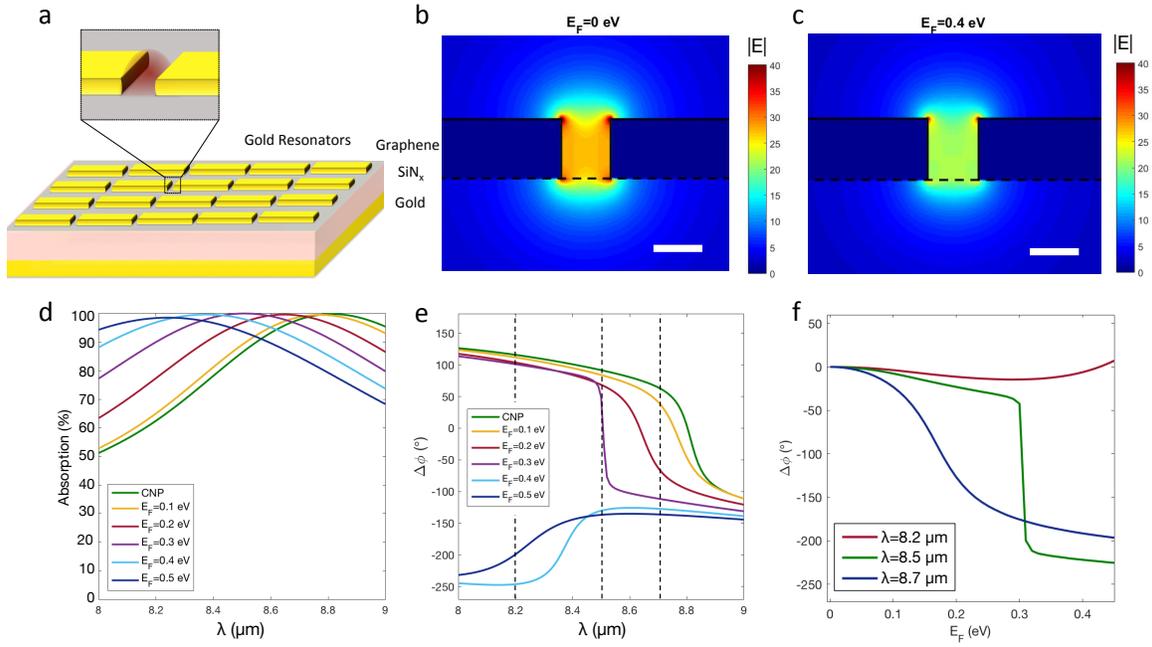

**Figure 1**. Tunable resonant gap-mode geometry. a) Schematic of graphene-tuned antenna arrays with field concentration at gap highlighted. Resonator dimensions: 1.2 μm length by 400 nm width by 60 nm height, spaced laterally by 50 nm. SiN$_x$ thickness 500 nm, Au reflector thickness 200 nm. b, c) Field profile in the antenna gap shows detuned resonance at different E$_F$ at a wavelength of 8.70 μm. Scale bar is 50 nm. d) Simulated tunable absorption for different graphene Fermi energies. e) Simulated tunable phase for different graphene Fermi energies. f) Phase modulation as a function of Fermi energy for three different wavelengths – 8.2 μm, 8.5 μm, 8.7 μm.

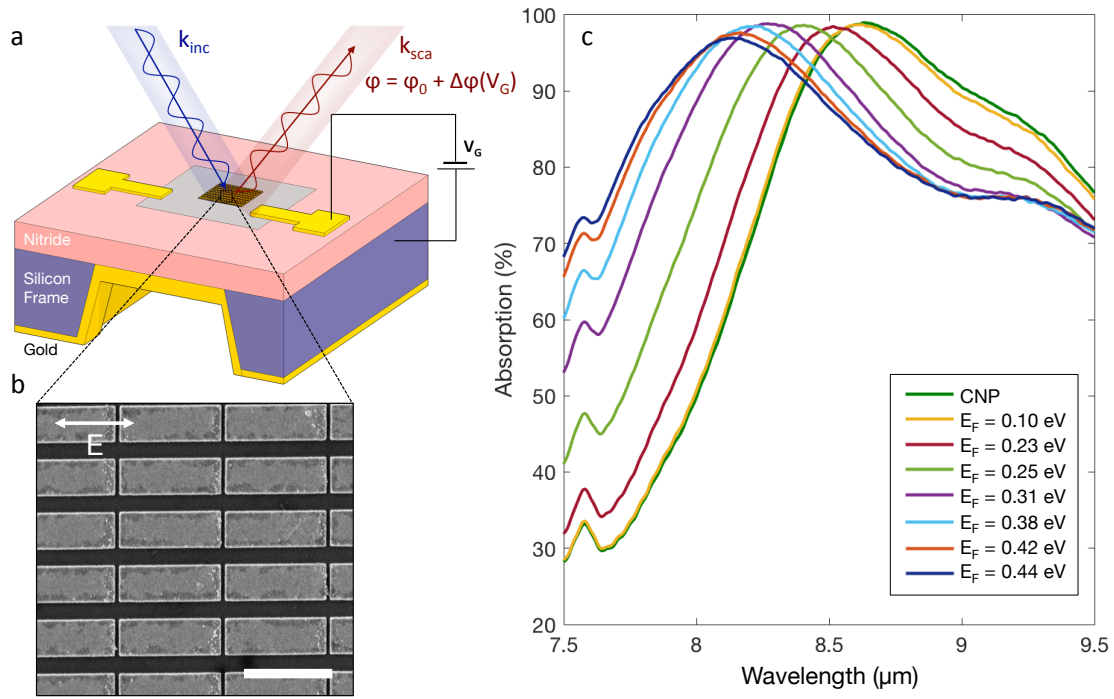

**Figure 2**. a) Schematic of a gate-tunable device for control of scattered phase. b) SEM image of gold resonators on graphene. Scale bar indicates 1 μm. c) Tunable absorption measured in FTIR at different gate voltages corresponding to indicated Fermi energies. A peak shift of 490 nm is measured.

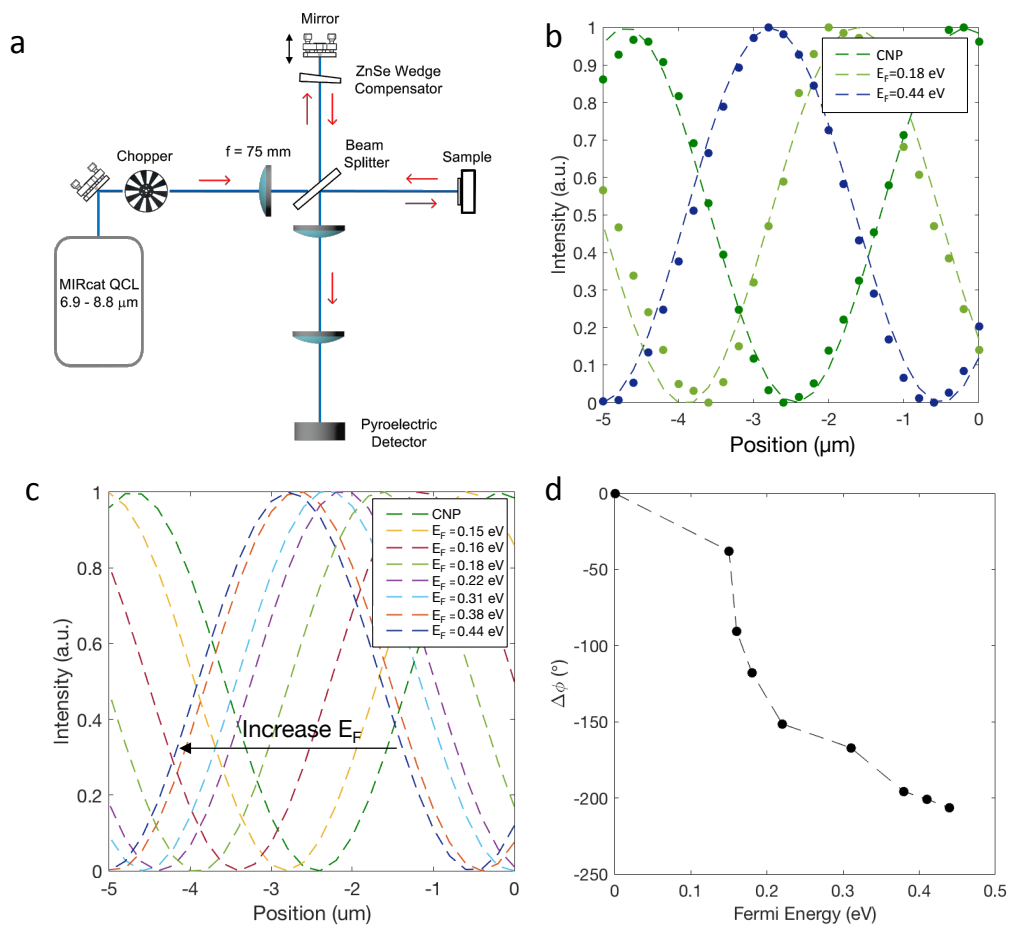

**Figure 3**. a) Schematic of a Michelson interferometer used to measure reflection phase modulation. b) Representative interferometer measurements for different Fermi energies with linear regression fits at a wavelength of 8.70 μm. c) Interferometry data fitted for all $E_F$ at 8.70 μm. d) Extracted phase modulation as a function of $E_F$ at 8.70 μm demonstrating 206° tuning.

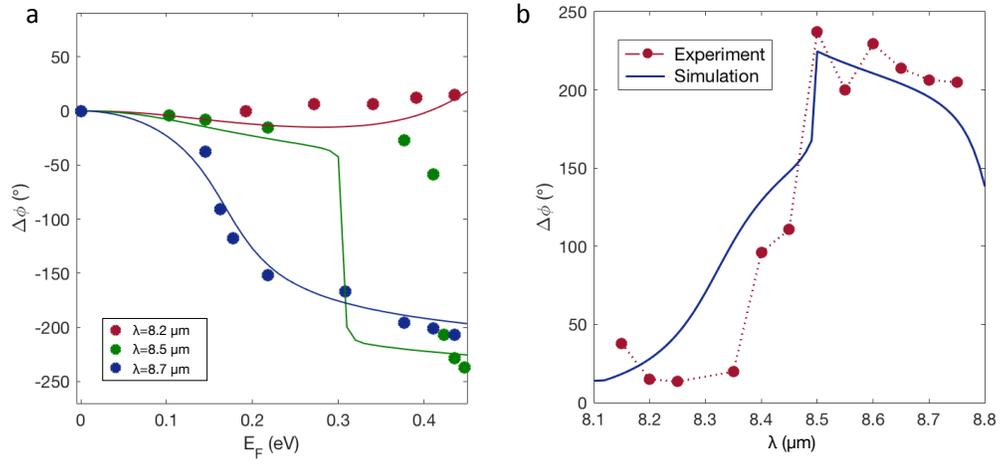

**Figure 4**. Demonstration of phase modulation over multiple wavelengths. a) Phase modulation at wavelengths of 8.20 μm, 8.50 μm, and 8.70 μm (circles – experiment, line – simulation). b) Maximum phase tuning achievable at wavelengths from 8.15 um to 8.75 um, simulation and experiment indicating up to 237° modulation.

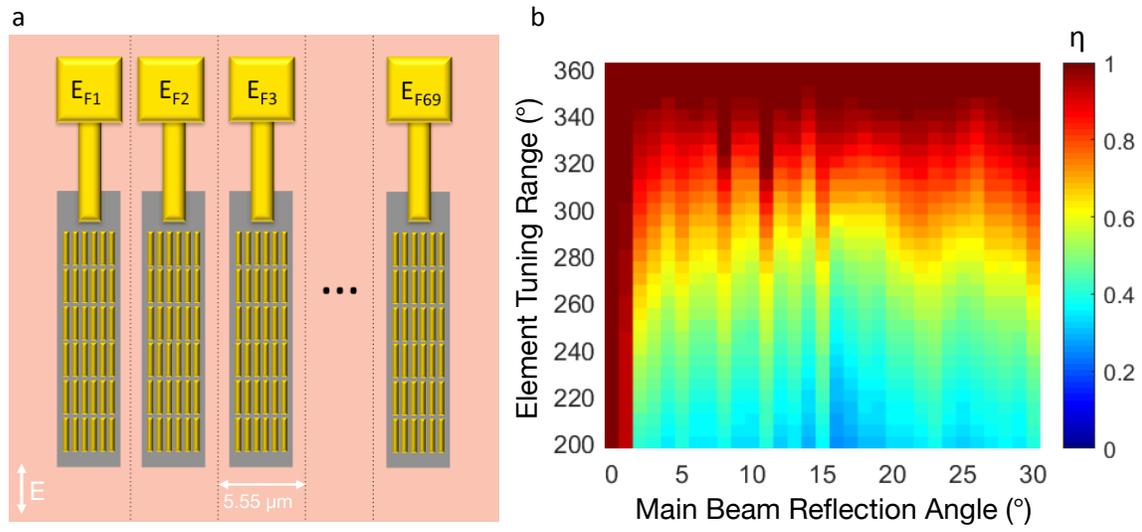

**Figure 5**. Calculation of proposed reconfigurable metasurface based on experimentally realized design. a) Schematic of beam steering device, where each of the 69 unit cells is assigned a different $E_F$. b) Relative steering efficiency, $\eta$, for an 69 element metasurface with a lattice spacing of 5.55 μm illuminated with a plane wave at 8.50 μm.